\documentclass{appolb}
\usepackage[dvips]{graphicx} 
\usepackage{epsfig,rotate} 
\usepackage{psfrag}
\usepackage{amsmath,amstext,amsbsy,amsopn,amsfonts,amssymb} 

\def \Im {\mathop{\rm Im}\nolimits}
\def \Re {\mathop{\rm Re}\nolimits}

\newcommand{\baxq}[2]{\mathbb{Q}({#1},{#2})} 
\def \bb  {\begin {thebibliography} }
\def \eb  {\end{thebibliography}}
\newcommand \bi [1] {\bibitem{#1}}

\psfrag{nu_h}[cc][bc]{ $\nu_h$}
\psfrag{-E_3}[cc][cc]{\small ${\epsilon_3}$}
\psfrag{-E_4}[cc][cc]{\small ${\epsilon_4}$}
\psfrag{-E_N/4}[cc][cc]{ ${\epsilon_N}$}
\psfrag{N}[cc][bc]{\small $N$}

\psfrag{Im(q3^(1/3))}[cc][cc]{\small 
\rotatebox{90}{$\Im[{q_3}^{1/3}]\mbox{ }$}}
\psfrag{Re(q3^(1/3))}[cc][bc]{\small $\Re[{q_3}^{1/3}]$}

\psfrag{Im(q4^(1/4))}[cc][cc]{\small
\rotatebox{90}{ $\Im[{q_4}^{1/4}]\mbox{ }$}}
\psfrag{Re(q4^(1/4))}[cc][bc]{\small $\Re[{q_4}^{1/4}]$}

\psfrag{Im[q3^(1/3)]}[cc][cc]{\small $\Im[{q_3}^{1/3}]$}
\psfrag{Re[q3^(1/3)]}[cc][bc]{\small $\Re[{ble}{q_3}^{1/3}]$}

\begin{document} 
   
\title{
%\hfill TPJU-11/2001\\ 
%~~\\ 
SPECTRUM OF THE MULTI-REGGEON COMPOUND STATES IN MULTI-COLOUR QCD 
%Spectrum the Multi-Reggeon Compound States in Multi-Colour QCD} 
\thanks{Presented at the X International Workshop on Deep Inelastic Scattering (DIS2002)}
\author{{\bf Jan Kota{\'n}ski}
\address{M. Smoluchowski Institute of Physics, Jagellonian University\\
Reymonta 4, 30-059 Krak{\'o}w, Poland}}}
                                 
\date{June 30, 2002} 
\maketitle 
 
\begin{abstract} 
We study the properties of the colour-singlet compound states
of regge\-ized gluons in multi-colour QCD. 
Applying the methods of integrable models, we calculate their
spectrum and discuss the application of the obtained results to 
high-energy asymptotics of the scattering amplitudes in perturbative QCD.
\end{abstract} 

The high-energy asymptotics of the scattering amplitude is governed by
the $t-$channel exchange of arbitrary number $N=2,3,...$ of reggeized
gluons described by the so-called ``fish-net'' or ``checker-board''
Feynman diagrams. The problem of resumming such diagrams
formulated in sixties \cite{FBAR}. 
%The Regge compound state problem, which allows us to describe
%scattering amplitude of two hadrons, was formulated in sixties.
The lowest non-trivial contribution for $N=2$ was calculated by
Balitski, Fadin, Kuraev and Lipatov in 1978 \cite{L2}, 
who derived and solved an equation for 
%the Pomeron intercept. 
intercept of the compound state of $N=2$ reggeized gluons -- the BFKL Pomeron.
The equation for three and more 
reggeized gluons
%Reggeons
was formulated in the 1980's by Bartels, Kwieci{\'n}ski, Prasza{\l}owicz
\cite{BKP} and Jaroszewicz \cite{jar}.
However, it took almost twenty years to obtain the solution for $N=3$, 
which corresponds to the QCD odderon \cite{JW,LN}. 
%During last year 
%S.\ \'E.\ Derkachov\footnote{Department of Mathematics, 
%St.-Petersburg Technology 
%Institute, St.-Petersburg, Russia}, G.\ P.\ Korchemsky\footnote{Laboratoire 
%de Physique Th\'eorique, Universit\'e de Paris XI,
% 91405 Orsay C\'edex, France}, 
%A.\ N.\ Manashov\footnote{Department d'ECM, Universitat de Barcelona, 
%08028 Barcelona, Spain} and the author of that paper 
%have calculated solutions for $N=2, \ldots, 8$ in the multi-colour limit
%\cite{KKM}.  
%This presentation is a result of the mentioned collaboration.
Finally, the solutions for higher $N=4,...,8$ and their generalizations
to arbitrary $N$ were obtained recently in the multi-colour limit in
series of papers \cite{DKM,KKM,DKKM} written in collaboration with
S.\ \'E.\ Derkachov\footnote{Department of Mathematics, 
St.-Petersburg Technology 
Institute, St.-Petersburg, Russia}, G.\ P.\ Korchemsky\footnote{Laboratoire 
de Physique Th\'eorique, Universit\'e de Paris XI,
 91405 Orsay C\'edex, France}, 
A.\ N.\ Manashov\footnote{Department d'ECM, Universitat de Barcelona, 
08028 Barcelona, Spain}. 

The leading contribution to the total elastic scattering amplitude
of two hadrons $(A,B)$ can be written in QCD in the Regge limit:
$s\rightarrow \infty$, $t=const$,
as a power series in a strong coupling constant 
$\overline{\alpha}_{s} = \alpha_s N_c / \pi$
\begin{equation}
{\cal A}(s,t) \sim - i \sum_{N=2}^{\infty} (i \overline{\alpha}_{s})^N
\frac{s^{1+\overline{\alpha}_{s} 
\epsilon_N}}
{\sqrt{\overline{\alpha}_{s} \ln s}}
\beta_A^{(N)}(t) \beta_B^{(N)}(t).
\end{equation}
%\begin{equation}
%A(s,t) \sim - i \sum_{N=2}^{\infty} (i \overline{\alpha}_{s})^N
%\frac{s^{1+\overline{\alpha}_{s} 
%\epsilon_N}}
%{\sqrt{\overline{\alpha}_{s} \ln s}}
%\beta_A^{(N)}(t) \beta_B^{(N)}(t).
%\end{equation}
Above, $N$ denotes the number of reggeized gluons, called Reggeons,
propagating in the $t$-channel and the residue factors, 
$\beta_{A(B)}^{(N)}$,
measure the overlap of 
%the Reggeon wave function 
the wave function of the compound state
of $N$ reggeized gluons
with the wave functions
of two scattered particles. 
%In order to calculate $\epsilon_N$,
%the energy of the ground state, one should solve the 
%Schr\"odinger (BKP) equation
The parameter
$\epsilon_N$ is defined as the maximal energy for the Schr\"odinger
(BKP) equation
\begin{equation}
{\cal H}_N \Psi \left( \left\{ 
\vec{z}_k \right\}\right)= \epsilon_N 
\Psi\left(\left\{ \vec{z}_k \right\}\right)
\label{Schr}
\end{equation}
where $\Psi \left( \left\{\vec{z}_k \right\}\right)$ is the Reggeon
wave function and $\vec{z}_k$ denotes two-dimentional transverse coordinates
of $k^{th}$ reggeized gluon. ${\cal H}_N$
is the effective QCD Hamiltonian describing pair-wise interaction between
$N$ reggeons.

In the multi-colour limit, this Hamiltonian simplifies
significantly \cite{L1,FK} leading to
\begin{equation}
{\cal H}_N =\sum_{k=0}^{N-1} H(\vec{z}_k,\vec{z}_{k+1})
\;\; \;\; \mbox{ where } \; \; \;\; \vec{z}_0 \equiv \vec{z}_N.
\label{Ham_N}
\end{equation}
It describes \cite{KKM} the nearest neighbour interaction of the Reggeons
and has a hidden cyclic and mirror permutational symmetry. 
Moreover, it possesses the set of the $(N-1)$ integrals of motion, which are 
the eigenvalues of conformal charges, $\hat{q}_k$ and 
$\hat{\overline{q}}_k$\footnote{{\em bar} does not denote complex              conjugation for which we use an {\em asterisk}.} \cite{DKM}, 
%for arbitrary number of Reggeons 
\begin{equation}
\left[{\cal H}_N, \hat{q}_n\right]
=\left[\hat{q}_n, \hat{q}_m\right]=
\left[{\cal H}_N, \hat{\overline{q}}_n\right]
=\left[\hat{\overline{q}}_n, \hat{\overline{q}}_m\right]
=0,\;\;\;\; n,m  = 2,\ldots,N.
\end{equation}
Thus, this system is completely integrable. The lowest integral
may be expressed in terms of the confromal $SL(2)$ weight of the state
$h$ as $q_2=-h(h-1)$.
The Hamiltonian (\ref{Ham_N}) is equivalent to the XXX Heisenberg spin magnet
\cite{DKM}.

%Our method uses the Baxter $Q$-operator. It has to commute with itself
%and with the integrals of motion
Our solution of the Schr\"odinger equation (\ref{Schr}) is
based on the method of the Baxter Q-operator \cite{DKM}. It relies on the
existence of the operator $Q(u,\bar u)$ depending on the pair of
complex spectral parameters $u$ and $\bar u$ and satisfying the 
following relations. It commutes with itself for different
values of the spectral parameters and with the integrals of motion
\begin{equation}
\left[\baxq{u}{\overline{u}},\baxq{v}{\overline{v}}\right]=
\left[\hat{t}_N(u,\left\{\hat{q}_n\right\}),\baxq{v}{\overline{v}}\right]=
\left[\hat{t}_N(\overline{u},\left\{\hat{\overline{q}}_n\right\}),
\baxq{v}{\overline{v}}\right]=0,
\end{equation}
where
\begin{equation}
\hat{t}_N(u,\left\{\hat{q}_n\right\})=2 u^N+
\hat{q}_2 u^{N-1}+\ldots+\hat{q}_N,
\end{equation}
and $u$, $v$ are two complex spectral parameters.
It also has to satisfy the Baxter equations
\begin{equation}
\begin{array}{ccl}
\hat{t}_N(u,\left\{\hat{q}_n\right\}) \baxq{u}{\overline{u}}&=& 
u^{N}\baxq{u+i}{\overline{u}}+ u^{N}\baxq{u-i}{\overline{u}} \\
\hat{t}_N(\overline{u},\left\{\hat{\overline{q}}_n\right\})\baxq{u}{\overline{u}}&=& 
(\overline{u}+i)^{N}\baxq{u}{\overline{u}+i}+
(\overline{u}-i)^{N}\baxq{u}{\overline{u}-i}.
\end{array}
\end{equation}
Furthermore, the $Q$-operator has prescribed analytical properties, 
i.e. known pole structure, and asymptotic behaviour at infinity.
The above conditons fix the $Q$-operator uniquely and allow us to quantize 
the integrals $q_k$ \cite{DKM,MW,DL,KP}.
It turns out that it is possible to express the Hamiltonian (\ref{Ham_N}) 
in terms of the Baxter $Q$-operator \cite{DKM}. Combining together the
solutions of the Baxter equations and the quantum conditions 
for $q_k$ with the Schr\"odinger equation (\ref{Schr}) 
we can calculate the energy spectrum.

\begin{figure}[ht]
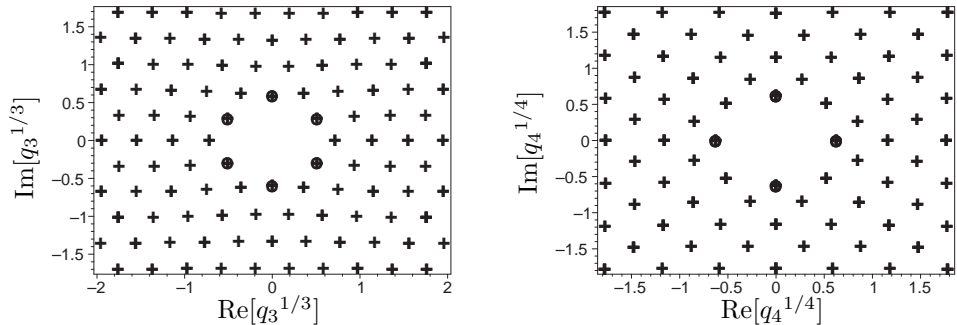

\vspace*{3mm}
\centerline{{\epsfysize4.0cm \epsfbox{n33.eps}}
\quad{\epsfysize4.0cm \epsfbox{n44.eps}}}
%\caption[]{Quantum values of the integrals of motion $q_N$ ($N=3, 4$)}
\caption[]{Quantized values of
the integrals of motion at $h=1/2$ for different number
of reggeons $N=3$ (left) and $N=4$ (right).}
\label{q3}
\end{figure}

For $N=3$ there exist two integrals of motion, $q_2$ and $q_3$.
%If we calculate cube root of $q_3$ at $h=1/2$ 
%we obtain symmetrical structure (plot \ref{q3} on the left) 
The quantized values of $q_3$ exhibit some
structure that can be seen on Figure \ref{q3} (left panel)
\cite{DKKM}. 
The circled crosses denote the ground states. 
They have the highest energy for fixed $N$.
%This spectrum may be approximated by following the
%equilateral triangle formula 
The spectrum of quantized $q_3$
at $N=3$ may be approximated by the following formula
\cite{WKB}
\begin{equation}
\left[{q_3}^{\mbox{\tiny approx}}(\ell_1,\ell_2)\right]^{1/3}=
\frac{\Gamma^3(2/3)}{2 \pi} 
\left(\frac{1}{2} \ell_1 + i \frac{\sqrt{3}}{2} \ell_2 \right)
\end{equation}
where $\ell_1$, $\ell_2 \in {\mathbb Z}$ and $\ell_1+\ell_2$ is even.
Similar structure is observed for $N=4$, $h=1/2$ and $q_3=0$ 
(plot 1 on the right)
\begin{equation}
\left[q_4^{\mbox{\tiny approx}}(\ell_1,\ell_2)\right]^{1/4}=
\frac{\Gamma^2(3/4)}{2 \sqrt{\pi}}
\left(\frac{1}{\sqrt{2}} \ell_1 + i \frac{1}{\sqrt{2}} \ell_2 \right).
\end{equation}
%Of course, there exist 
%other states with $q_3\ne 0$ but they have lower values of energy.
All these states have continuation in $\nu_h=\Im(h)$, so, the eigenvalues
of $\left\{\hat{q}_k\right\}$ form linear trajectories in the full ${q_k}$ space \cite{braun,PR}.
The spectrum of $q_N$ has similar structure for higher $N$.
%The spectrum of $\hat{q}_k$'s
%possesses for higher $N$'s the similar structure 
%but in more-dimensional space.

\begin{table}[h]
{\small
\begin{center}
\begin{tabular}{|c||c|c|c|c|c|c||c|}
\hline
 $N$ & $ i{q_3}$ &  ${q_4}$ & $i{q_5}$ &${q_6}$ &$i{q_7}$ &
${q_8}$ & ${\epsilon_N}$ 
\\
\hline
$2$ &      &     &   &    &&       & 2.7726 
\\
${3}$ & .20526 &   &     &    &&       & -.2472
\\
${4}$ &   0   & .15359 &    &   &&   & \phantom{-}.6742 
\\
${5}$ & .26768 & .03945 & .06024 &    &&    & -.1275
\\
${6}$ & 0 & .28182 & 0  & .07049 &&& \phantom{-}.3946 
\\
$7$ &.31307 & .07099 &.12846 & .00849 & .01950 & & -.0814 
\\
$8$ & 0 & .39117 & 0 & .17908 & 0 & .03043 & .2810
\\
\hline
\end{tabular}
\end{center}}
\caption{Quantum numbers $q_k$ and energy, $\epsilon_N(\{q_k\})$, 
of the $N$-Reggeons states
in the multi-colour QCD for $h=1/2$.}
\label{tabqe}
\end{table}

\begin{figure}[ht]
\vspace*{3mm}
\centerline{{\epsfysize 4.0cm \epsfbox{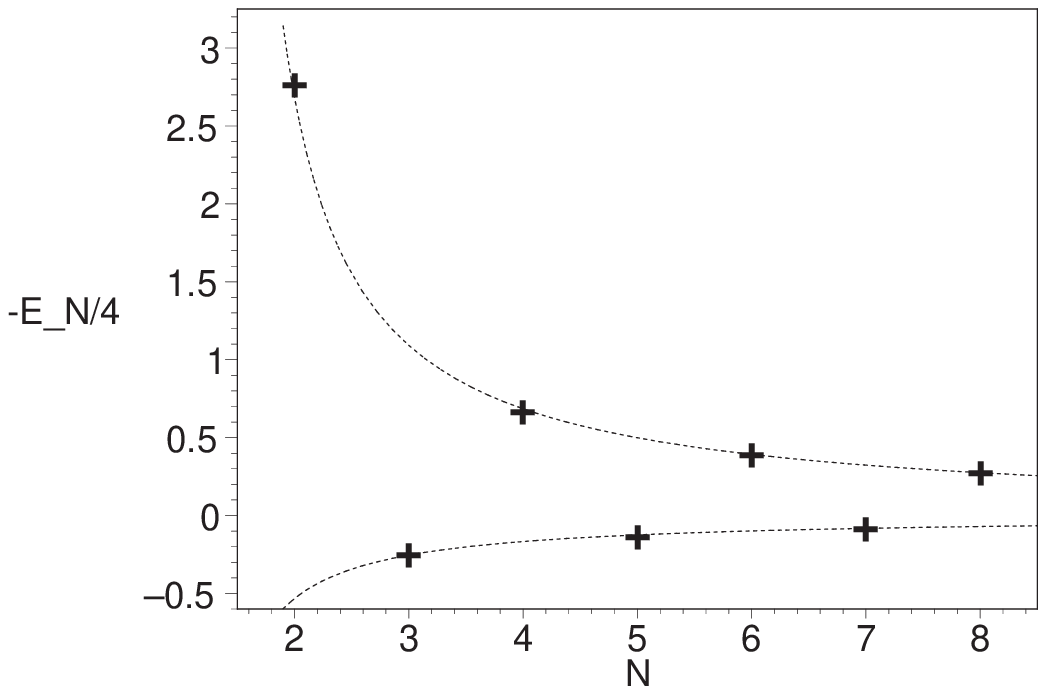}}
\quad{\epsfysize 4.0cm \epsfbox{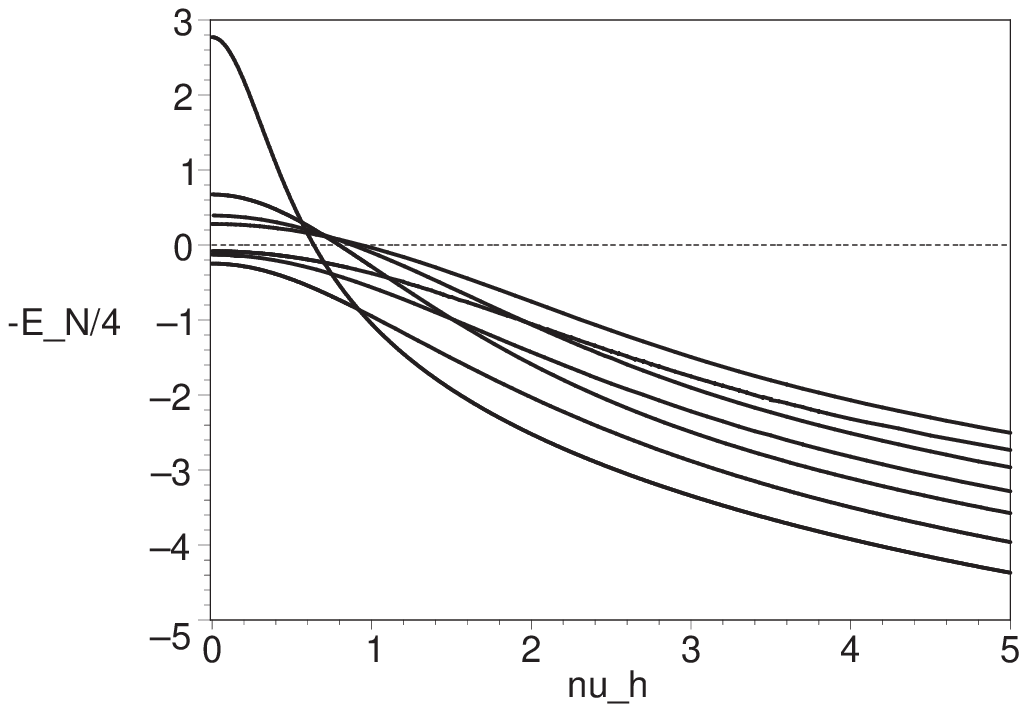}}}
\caption[]{\small The dependence of the ground state energy, 
$\epsilon_N$, on the number of
particles $N$ and on the $\nu_h$. }
\label{en}
\end{figure}

For the ground states (tab. \ref{tabqe}), the integrals of motion
are either purely real or purely imaginary. Moreover, the odd integrals
are equal to zero for even $N$'s. 
%These numbers agree with the previously 
%published ones 
The numbers at $N=2$ and $N=3$ agree with the previously published
ones \cite{L2,JW,KKM,KP,PR}, while the results for higher $N$ are new.
The energy is positive for
even $N$'s and negative for odd $N$'s and it is shown in a plot \ref{en}.
The contribution of these states to the scattering amplitude increases 
with $s$ for the positive energy $\epsilon_N$ and decreases
for the negative energy $\epsilon_N$.
The exact values of the energy are denoted by crosses on the left panel
of fig. \ref{en}. 
The upper and the lower curves stand for the functions 
$1.8402/(N-1.3143)$ and $-2.0594/(N-1.0877)$, respectively.

The right panel in the figure \ref{en}
shows the dependence of the energy $\epsilon(\nu_h)$
along the ground state trajectory for different number of particles 
$2\le N \le 8$. 
These functions are symmetrical in $\nu_h$. The maximum energy
is in $\nu_h=0$.
At large $\nu_h$, $\epsilon_8>...>\epsilon_3>\epsilon_2$.

Summerizing we found the spectrum of the multi-Reggeon compound states in QCD. 
 In the Pomeron sector (even $N$) the intercept of states is bigger
than $1$ but smaller than the intercept of the BFKL Pomeron:
$\alpha_2 > \alpha_4 > \ldots > 1$.
In the odderon sector (odd N) the intercept of the states is 
smaller than $1$ but it increases with $N$:
$\alpha_3 < \alpha_5 < \ldots < 1$.
%In the future, it is possible to calculate the scattering amplitude
%with higher precision.
%One can also try to find some analytical explanations of
%our numerical results.

Recently Lipatov and de Vega \cite{VL} found another set of
the solutions for $N=3, 4$ 
%compound states 
which differ from our expressions. This discrepancy requires
further studies.
% solutions for $N=2,\ldots,4$ 
%gluon exchanges based on different quantization conditions.

%\bibliographystyle{plain} 

\end{document}